\documentclass[aps,prb,twocolumn,superscriptaddress]{revtex4}
\usepackage{graphicx}
\usepackage{amsmath}
\usepackage{amssymb}
\usepackage{epstopdf}

\begin{document}

\title{Short-range structure of proton conducting perovskite BaIn$_{x}$Zr$_{1-x}$O$_{3-x/2}$ ($x=0-0.75$)
}
\author{Maths Karlsson}
\affiliation{Department of Applied Physics,
             Chalmers University of Technology,
             SE-412 96 G{\"o}teborg, Sweden}
\author{Aleksandar~Matic}
\affiliation{Department of Applied Physics,
            Chalmers University of Technology,
            SE-412 96 G{\"o}teborg, Sweden}
\author{Christopher S.~Knee}
\affiliation{Department of Chemistry,
            G{\"o}teborg University,
            SE-412 96 G{\"o}teborg, Sweden}
\author{Istaq~Ahmed}
\affiliation{Department of Chemical and Biological Engineering,
	   Chalmers University of Technology,
            SE-412 96 G{\"o}teborg, Sweden}
\author{Sten~Eriksson}
\affiliation{Department of Chemical and Biological Engineering,
	   Chalmers University of Technology,
            SE-412 96 G{\"o}teborg, Sweden}
\author{Lars B{\"o}rjesson}
\affiliation{Department of Applied Physics,
             Chalmers University of Technology,
             SE-412 96 G{\"o}teborg, Sweden}

\date{\today}
\begin{abstract}
In a systematic study we investigate the effect of dopant level and hydration on the short-range structure of the proton conducting perovskite-type oxide BaIn$_{x}$Zr$_{1-x}$O$_{3-x/2}$ ($x=0-0.75$), using infrared and Raman spectroscopy. The results show that doping leads to significant local distortions of the average cubic structure of these materials. By increasing the In concentration from $x=0$ to $x=0.75$ new bands appear and grow in intensity in both the IR and Raman spectra, showing that the local distortions become successively more and more pronounced. The structural distortions are largely uncorrelated to the presence of oxygen vacancies, but instead are mainly driven by the size difference between the In$^{3+}$ and Zr$^{4+}$ ions, which leads to displacements of the cations and to tilting of the (In/Zr)O$_{6}$ octahedra. Based on our results, we suggest that there is a threshold between $x=0.10$ and $x=0.25$ where the local structural distortions propagate throughout the whole perovskite structure. Comparison of our spectroscopic data with the proton conductivity reported for the same materials indicates that the presence of extended structural distortions are favorable for fast proton transport.

\end{abstract}

\maketitle
\section{Introduction}

Perovskite-type oxides, ABO$_{3}$, (hereafter shortly denoted perovskites) have attracted considerable attention because of their technological use and academic interest. 
Proton conduction in this class of materials was first discovered in 1981 by Iwahara \textit{et al.}\cite{IWA81} and has been widely investigated, especially since these materials can be used in various electrochemical applications such as fuel cells, gas sensors and hydrogen pumps.\cite{IWA92}
Protons are introduced into the perovskite structure by the formation of an oxygen-deficient structure and subsequent hydration. The oxygen vacancies are usually formed as a charge-compensating effect by doping the ideal perovskite A$^{2+}$B$^{4+}$O$^{2-}_{3}$ with lower-valent B$^{\prime}$$^{3+}$ ions. The general formula of the acceptor-doped perovskite may then be written as A$^{2+}$B$^{4+}_{1-x}$B$^{\prime}$$^{3+}_{x}$O$^{2-}_{3-\delta}$, where $\delta$ refers to the oxygen vacancies. Hydration can be performed by annealing in a humid atmosphere at elevated temperatures. During this procedure, the water molecules dissociate into hydroxide ions which fill the oxygen vacancies while each of the remaining protons forms a covalent bond with one lattice oxygen.\cite{KRE03}

Supported by several independent neutron scattering experiments\cite{HEM95, MAT96, PIO97} and simulations\cite{MUN97, KRE98, MUN96, SHI97} it is widely suggested that the proton conduction mechanism in these materials occurs through two elementary steps; i) proton transfer between neighboring oxygens via hydrogen bonds, and ii) rotational motion of the hydroxyl group in between such transfers.\cite{KRE03} 
Thus, the local structure around the proton is a key factor for the proton transport in these materials.

The structure of perovskites is routinely characterized by X-ray and neutron diffraction. 
However, these techniques give information only about the long-range average structure of materials.
Thus, the presence of possible local structural distortions cannot be directly revealed from such measurements. 
This means that even though certain perovskites are referred to as cubic, their structures on a short-range scale ($<$50~{\AA}) may differ from the cubic periodicity.
It is, therefore, quite surprising that there are only a few experimental reports on the short-range structure of proton conducting perovskites.
To our knowledge, these investigations, which indeed have provided some important information, are limited to X-ray absorption fine structure (XAFS) investigations of doped perovskites based on SrCeO$_{3}$\cite{ARI99}, SrZrO$_{3}$\cite{KAM01}, BaCeO$_{3}$\cite{MAT97,WU05,LON06}, and CaZrO$_{3}$\cite{DAV00,ISL01_2}.
However, no investigations have focused on the effect of dopant level and the role of oxygen vacancies on the short-range structure and its direct relationship to the proton conductivity.
In order to understand the details of the proton transport in hydrated perovskites a thorough knowledge about the short-range structure is essential.

\begin{figure}
\includegraphics[width=0.45\textwidth]{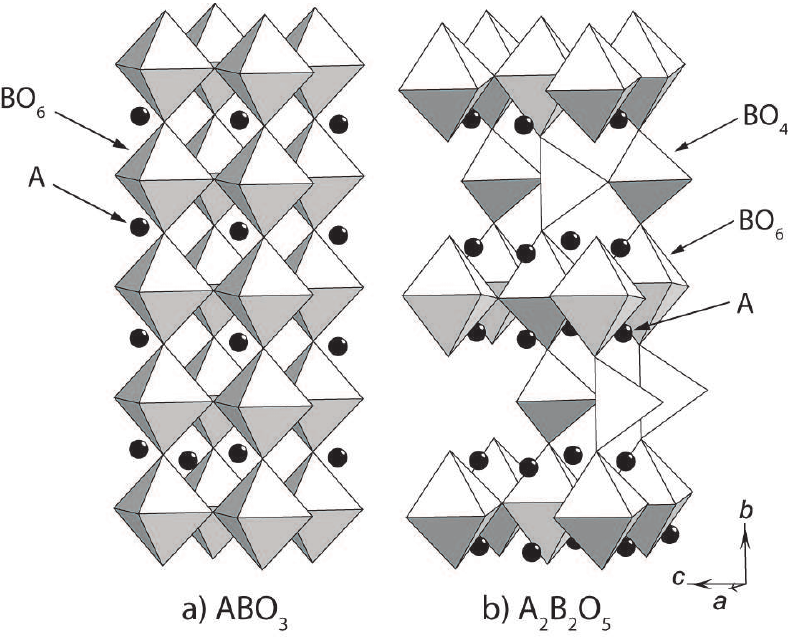}
\caption{\label{structure} Schematic picture of the ideal a) cubic perovskite ABO$_{3}$ and b) brownmillerite A$_{2}$B$_{2}$O$_{5}$ structures.}
\end{figure}

In this work we systematically investigate the short-range structure of dry and hydrated samples of the proton conducting perovskite BaIn$_{x}$Zr$_{1-x}$O$_{3-x/2}$ ($x~=~0~-~0.75$), and, for comparison, the brownmillerite end member Ba$_{2}$In$_{2}$O$_{5}$ ($x=1$).
The experiments are performed using infrared (IR) and Raman spectroscopy, which are two well suited techniques for these investigations.\cite{YE98, SIN98}  
In particular, these techniques give information about the oxygen phonons, a key factor that affects the rate of both the proton transfer and the rotational step in the proton conduction mechanism.\cite{MUN96}
From X-ray and neutron diffraction, the structure of BaIn$_{x}$Zr$_{1-x}$O$_{3-x/2}$ is found to be cubic (space group $Pm$$\bar{3}$$m$) with lattice parameter $a_{\rm{perov}}=$~4.19--4.25~{\AA} for $0~\leqslant$~$x$~$\lesssim~0.8$, and brownmillerite (space group $Icmm$) for $0.8\lesssim$~$x$~$\leqslant1$.\cite{BER02, AHM06_2, AHM06_3, GOO90, MAN93} 
The brownmillerite Ba$_{2}$In$_{2}$O$_{5}$ structure is referred to as an orthorhombic superstructure of the cubic perovskite structure with $a_{\rm{brownm}}\sim\sqrt{2}a_{\rm{perov}}$, $b_{\rm{brownm}}\sim4a_{\rm{perov}}$ and $c_{\rm{brownm}}\sim\sqrt{2}a_{\rm{perov}}$. 
A comparison of the cubic perovskite and brownmillerite structures is shown in Fig.~\ref{structure} from which it is evident that the presence of oxygen vacancies leads to tetrahedral coordination for half of the In$^{3+}$ ions in Ba$_{2}$In$_{2}$O$_{5}$. Furthermore, correlated tilts of the InO$_{6}$ octahedra are also present in the brownmillerite structure running along the $b$-direction of the unit cell.

We have in a previous work on the hydrated BaIn$_{x}$Zr$_{1-x}$O$_{3-x/2}$ system shown that the presence of dopant atoms and oxygen vacancies in the structure influences the vibrational dynamics of the protons.\cite{KAR05} 
However, the relationship between the concentration of dopant atoms and oxygen vacancies to the actual short-range structure of these materials has not been clarified.
In the present work we show that the short-range structure of BaIn$_{x}$Zr$_{1-x}$O$_{3-x/2}$ ($x~=~0~-~0.75$) becomes successively more and more distorted from the ideal cubic symmetry upon increasing In concentration, and that these distortions are largely uncorrelated to the presence of oxygen vacancies in the structure but instead directly driven by the dopant atoms.

\section{Experimental Section}
\subsection{Sample preparation}

Samples of  BaIn$_{x}$Zr$_{1-x}$O$_{3-x/2}$ ($x~=0-1$) were prepared by mixing stoichiometric amounts of BaCO$_{3}$, ZrO$_{2}$ and In$_{2}$O$_{3}$. The oxides were heated to 800$^{\circ}$C overnight to remove moisture prior to weighing. To ensure thorough mixing, ethanol (99.5 {\%}) was added during the milling procedure, which was performed manually using an agate mortar and a pestle. The finely ground mixtures were fired at 1000$^{\circ}$C for 8 hours and subsequently ground and pelletized using a 13 mm diameter die under a pressure of 8 tons. The pellets were sintered at 1200$^{\circ}$C in air for 72 hours. After sintering, the pellets were reground, compacted and refired at 1325$^{\circ}$C for 48~h. Finally, the pellets were finely reground to powders. 
X-ray diffraction revealed a cubic symmetry for $x=0-0.75$ while the $x=1$ material could be indexed to a brownmillerite structure, in accordance with data reported elsewhere.\cite{BER02, AHM06_2, AHM06_3, GOO90, MAN93}
The hydration was performed by annealing the powder samples at 275--300$^{\circ}$C under a flow (12 ml/min) of Ar saturated with water vapor at 76$^{\circ}$C for 10 days. 
Dehydrated (dry) samples were prepared by annealing the as-prepared samples under vacuum ($\sim$2$\cdot$10$^{-6}$ mbar) at 950$^{\circ}$C overnight to remove as many protons as possible.

\subsection{Infrared spectroscopy}
Three different experimental setups were used to measure the IR spectra in the range 75--1000 cm$^{-1}$. The 75--450 cm$^{-1}$ part is the merged spectra using a Bruker IFS 66v/s FT-IR spectrometer equipped with a DTGS (Deuterated TriGlycerine Sulfate) detector and exchangeable beam splitters; Mylar 12 (75--220 cm$^{-1}$) and Mylar 3.5 ( 220--450 cm$^{-1}$). 
The 400--1000 cm$^{-1}$ part was measured in an inert atmosphere with a Bruker VECTOR 22 FT-IR spectrometer, equipped with a KBr beam splitter and a DTGS detector. 
The experiments described above were all performed in a transmittance setup with the samples dispersed to 5 weight percent in a 0.1 g polyethylene pellet (75--450 cm$^{-1}$ part) and KBr-pellet (400--1000 cm$^{-1}$ part), respectively. 
As the far-IR (75--450 cm$^{-1}$) and the mid-IR (400--1000 cm$^{-1}$) spectra were measured with the samples dispersed in different pellets, these spectra have not been merged. 
A load of 7 ton was applied to the powder mixtures in order to obtain the final $\O$13 mm pellets. The sample preparations for all experiments except the pellet pressing were performed in an inert atmosphere.

\subsection{Raman spectroscopy}
The Raman spectroscopy experiments were performed in double subtractive mode on a DILOR XY800 tripple-grating spectrometer equipped with a liquid-nitrogen cooled CCD detector. The measurements were performed in back-scattering geometry with a spot size around 30 $\mu$m in diameter. The samples were kept in an inert atmosphere in air tight cells. The 488 nm line from an Ar-ion laser was used for excitation. 

\section{Vibrational spectra of cubic perovskites}
\label{theory}
Perfectly cubic perovskites ABO$_{3}$ (space group $Pm$$\bar{3}$$m$) exhibit four types of optically active vibrations, 
\begin{equation}
\label{Representation}
\Gamma~= ~3f_{\rm{1u}}(IR)+f_{\rm{2u}}(S),
\end{equation}
of which three are IR-active (IR), one is silent (S) and none are Raman-active.\cite{LAS57}
The three IR-active vibrations, A-(BO$_{6}$) and B-O stretches, and O-B-O bends, are each triply degenerate since three equivalent axes exist.\cite{LAS57}
The A-(BO$_{6}$) vibration can be treated by considering the BO$_{6}$ octahedron as a single atom situated at the B site. 
In orthorhombic perovskites the degeneracy is completely removed and each band is tripled while in tetragonal and rhombohedral perovskites the degeneracy is partially removed and each band is split into two.
Furthermore, the normally silent mode $f_{\rm{2u}}$, related to a torsional motion of the unit cell, can become IR-active as a result of structural distortions of the ideal cubic symmetry of the perovskite.\cite{LAS57} 
Specifically, this mode becomes active when the B cation is not at a center of symmetry, a condition which is realized if the lattice is sheared.\cite{PER65}
Thus, the number of bands in the IR spectrum is directly related to the structure and possible local distortions of the perovskite. 
However, in perovskites with slight distortions of the cubic symmetry the degeneracy may not be split into well-defined bands. Instead, broad bands and shoulders may be observed. 
The broadening results from a distribution of several slightly different configurations of the vibrating units in the material, while shoulders arise from configurations of more distinctly different symmetry.
The fact that a perfectly cubic perovskite exhibits no Raman-active vibrations means that its Raman spectrum is totally flat. 
Any perturbation of the cubic perovskite structure can, however, make vibrations Raman-active.
In addition, even in non-distorted systems, combinations of phonons at various points in the Brilluoin zone, so called second order scattering, can contribute to the Raman spectrum.\cite{SCH66,RAN05} 

The frequencies of the host-lattice vibrations of perovskites typically fall into the range of 50 to 900 cm$^{-1}$, see \textit{e.g.} Refs.\cite{MAC04,COU74,RAN05,COL01}
A characteristic signature of protons in the structure are the O-H stretch vibrations, which usually have frequencies between 2500 and 3500 cm$^{-1}$,\cite{KAR05,KRE99} and thus these modes do not impinge on the bands related to the host-lattice vibrations of the perovskite.
Therefore, the IR and Raman spectra up to 1000 cm$^{-1}$ are expected to be related to the host-lattice vibrations of the perovskite, and hence reflect the short-range structure of the material under investigation.

\section{Results}
\subsection{Infrared spectroscopy}
\begin{figure*}
\includegraphics[width=0.8\textwidth]{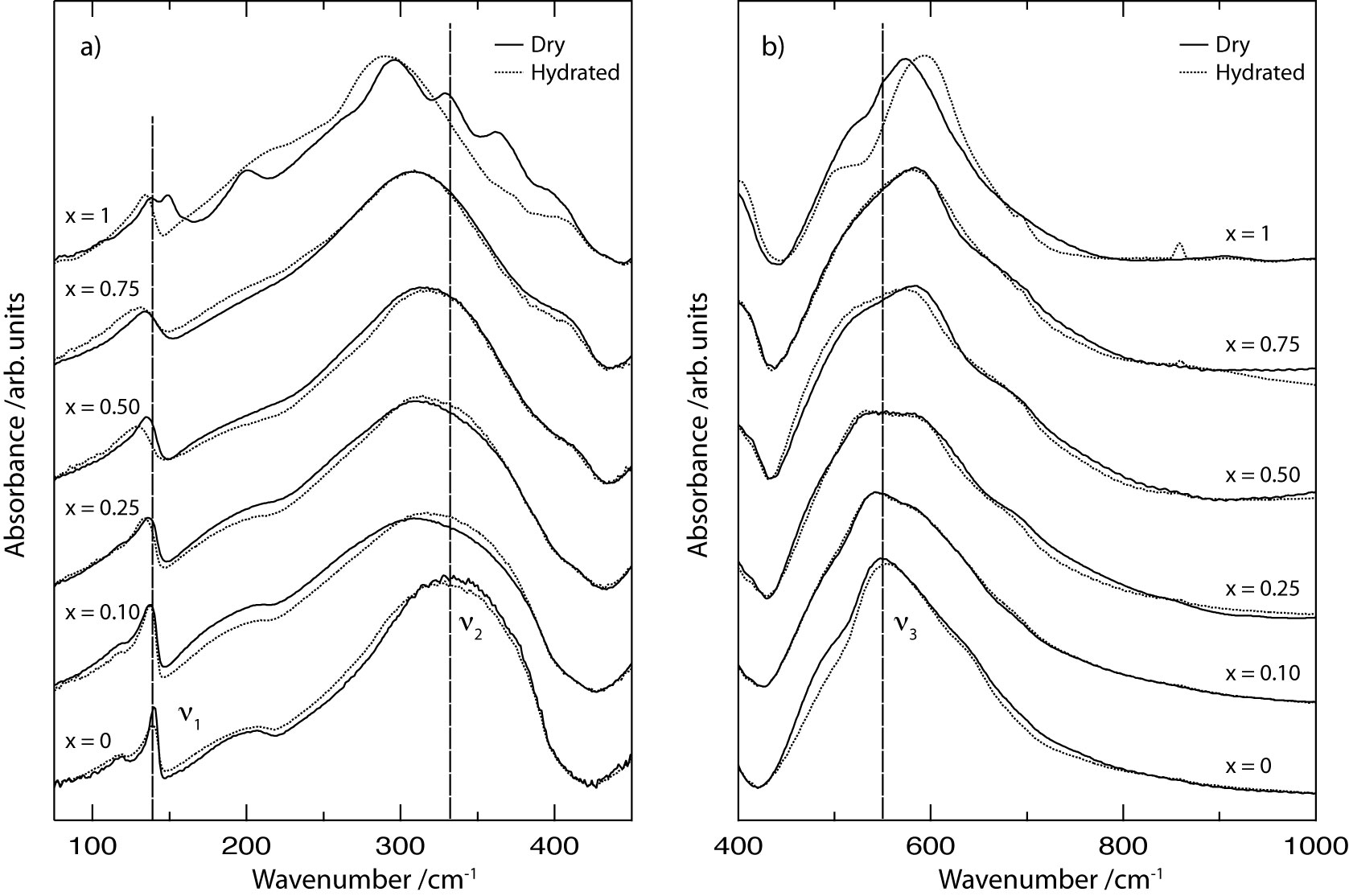}
\caption{\label{IR} Infrared spectra of dry and hydrated samples of BaIn$_{x}$Zr$_{1-x}$O$_{3-x/2}$ ($x$ = 0--1). The spectra have been vertically offset for clarity. The spectra show three strong bands ($\nu_{1}$, $\nu_{2}$ and $\nu_{3}$) attributed to Ba-[ZrO$_{6}$] stretches, O-Zr-O bends, and Zr-O stretches, respectively.
}
\end{figure*}
Figure~\ref{IR} shows the IR spectra of dry and hydrated samples of BaZr$_{1-x}$In$_{x}$O$_{x/2}$, with $x=0$, 0.10, 0.25, 0.50, 0.75 and 1, in the range of 50--1000 cm$^{-1}$.
It is immediately apparent that, except for $x=1$, the spectra of the dry and the hydrated samples are very similar.

In the spectrum of the undoped perovskite, BaZrO$_{3}$, we find three strong bands at around 135, 330 and 550 cm$^{-1}$ while a weaker band is located at approximately 200 cm$^{-1}$. 
The three strong bands at 135, 330 and 550 cm$^{-1}$ clearly lie in the range expected for cubic perovskites and can be assigned to the Ba-[ZrO$_{6}$] cation stretch ($\nu_{1}$), the O-Zr-O bend ($\nu_{2}$) and the Zr-O stretch vibrations ($\nu_{3}$).\cite{LAS57}
One should note the broad and asymmetric nature of the O-Zr-O bend and Zr-O stretch bands.
In addition, we observe shoulders on both sides of the Zr-O stretch band, $\nu_{3}$.
The weaker band at around 200 cm$^{-1}$ can, most likely, be assigned to the $f_{2u}$ mode, $i.e.$ a torsional motion of the oxygen octahedra.
This assignment is based on calculations, which have shown that it should lie in between the $\nu_{1}$ and $\nu_{2}$ modes,\cite{NAK67} as found in our spectra.

Regarding the spectra of the doped perovskites ($x~=~0.10-0.75$) in Fig.~\ref{IR} we observe several gradual changes with increasing In concentration. 
More specifically, each of the $\nu_{1}$ and $\nu_{2}$ bands broadens and shifts slightly towards lower frequencies while the $\nu_{3}$ band significantly changes shape. 
For the $\nu_{3}$ band, the high-frequency side becomes more pronounced and a weak shoulder at $\sim$680 cm$^{-1}$ can be discerned.
In addition, a new band at 415 cm$^{-1}$ can be resolved and this band gradually increases in intensity with the In concentration, see Fig.~\ref{IR}(a).

On further increase of the In concentration up to $x=1$, the change of the spectrum is particularly noticeable. 
For this composition, for which the structure is brownmillerite, the spectrum exhibits more bands than the spectra of the perovskites ($x\leqslant$~0.75), in particular in the region between 100 and 400 cm$^{-1}$.
The presence of more bands in the $x=1$ spectrum is a direct consequence of the lower symmetry of this material.
The $x=1$ spectrum is in agreement with that already reported for the same material.\cite{TEN05}

As noted above, it is only for the $x=1$ material that we can observe a significant spectral change upon hydration.
For the other investigated materials, the spectra of the dry and hydrated counterparts are essentially the same.
Interestingly, apart from the fact that the $\nu_{2}$ band is slightly down-shifted and the shoulder at the high-frequency side of the $\nu_{3}$ band is more intense, the spectrum of the hydrated $x~=~1$ material is very similar to those for $x~=~0.75$.

\subsection{Raman spectroscopy}

\begin{figure}
\includegraphics[width=0.4\textwidth]{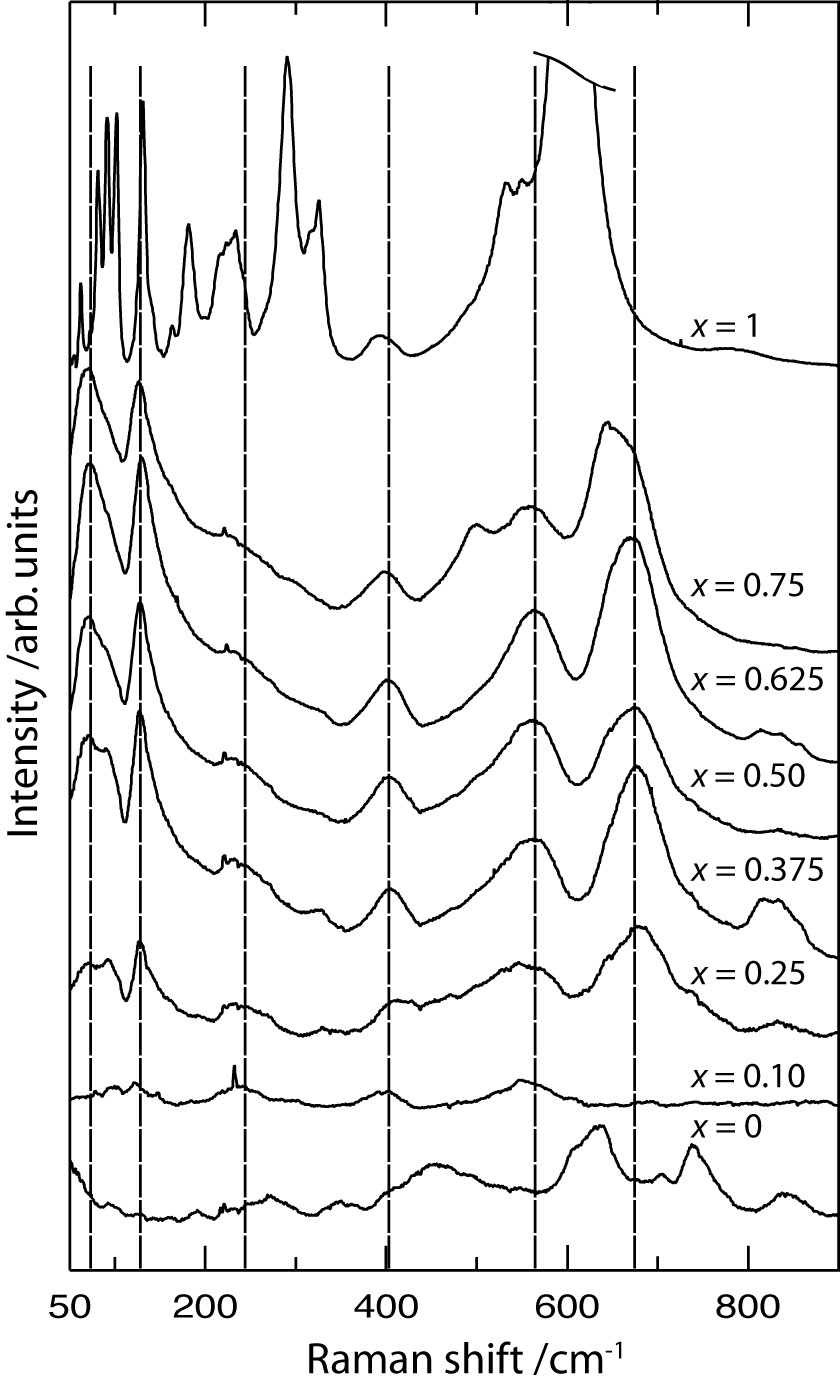}
\caption{\label{Ramandry} Raman spectra of dry samples of BaIn$_{x}$Zr$_{1-x}$O$_{3-x/2}$ (\textit{x}=0--1). For clarity all spectra have been vertically offset and the strong peak at $\sim$600 cm$^{-1}$ in the $x=1$ spectrum has been cut at approximately one fourth of its peak intensity.
}
\end{figure}
\begin{figure}
\includegraphics[width=0.4\textwidth]{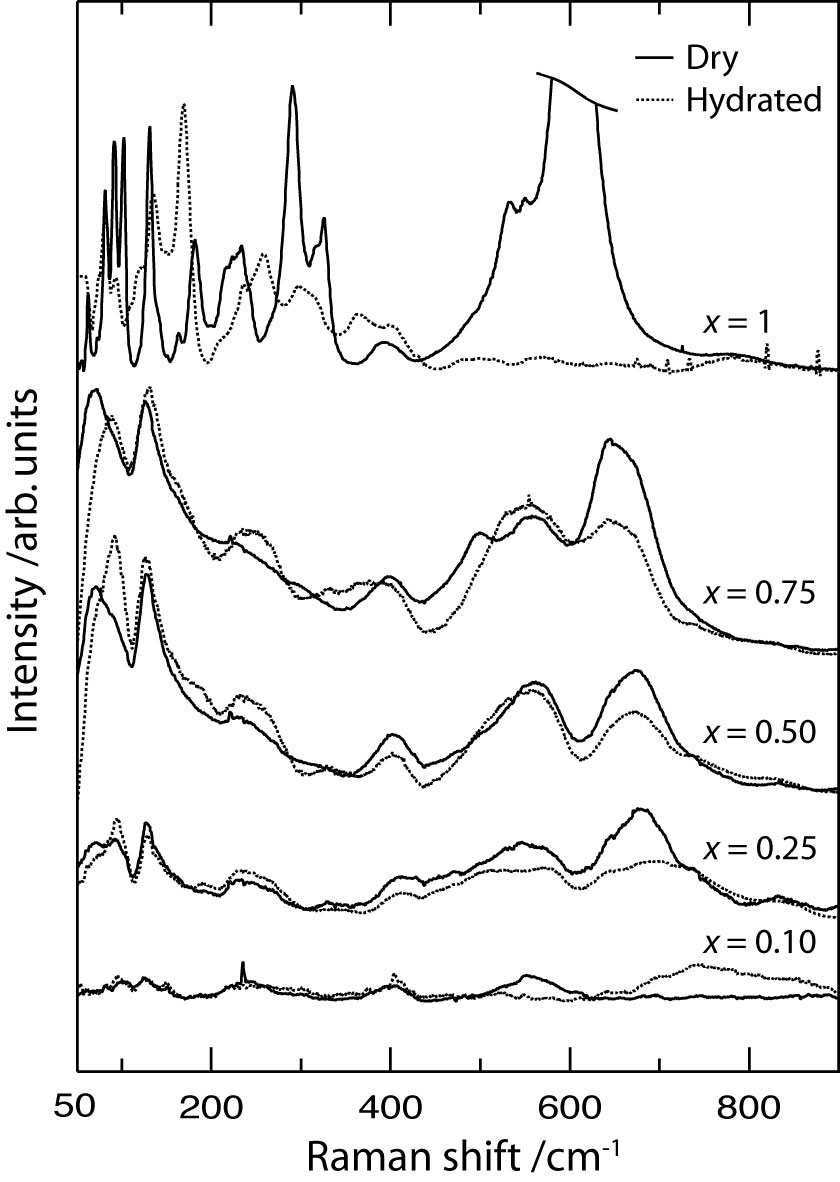}
\caption{\label{Ramanhydrated} Raman spectra of dry and hydrated samples of BaIn$_{x}$Zr$_{1-x}$O$_{3-x/2}$ (\textit{x}=0--1). For clarity all spectra have been vertically offset and the strong peak at $\sim$600 cm$^{-1}$ in the $x=1$ spectrum has been cut at approximately one fourth of its peak intensity.
}
\end{figure}

Figure~\ref{Ramandry} shows the Raman spectra of dry samples of BaIn$_{x}$Zr$_{1-x}$O$_{3-x/2}$ in the range 50--900 cm$^{-1}$.
First we observe that the spectrum of BaZrO$_{3}$ agrees well with those spectra already reported for the same material.\cite{CHA99,GEN97} 
For $x=0.10$ we observe that the Raman spectrum shows less distinct features than for $x=0$.
For $x=0.25$, we find several strong bands in the spectrum, and by further increasing the In concentration up to $x=0.75$ these bands become successively stronger. 

In the low-frequency part of the spectrum, two regions of bands, at $\sim$70--90 and at $\sim$130 cm$^{-1}$ are found for $x~\geqslant$~0.25.
The 70--90 cm$^{-1}$ region can be divided into two overlapping components at around 70 and 90 cm$^{-1}$, respectively. 
The particularly low frequency of these bands suggests that they are related to vibrations mainly involving Ba$^{2+}$ ions, which are the heaviest ions in these compounds and therefore expected to vibrate with the lowest frequency.
Furthermore, a gradual intensity increase of the 70 cm$^{-1}$ band and intensity decrease of the 90 cm$^{-1}$ band with dopant level suggest that these two bands also involve the In$^{3+}$ and Zr$^{4+}$ ions. 
Based on calculations\cite{GEN97} low-frequency vibrations, 90--120 cm$^{-1}$, in the Raman spectra of perovskites have previously been assigned to octahedron modes with a strong librational character coupled to vibrations of the A site ion.\cite{COL01}
On these grounds we suggest that they are related to deformational motions of Ba-[(Zr/In)O$_{6}$] units. 
The 130 cm$^{-1}$ band is most likely associated to the Ba-[(Zr/In)O$_{6}$] stretch vibration as this mode is found at nearly the same frequency ($\sim$135 cm$^{-1}$) in the corresponding IR spectra, as shown in Fig.~\ref{IR}.
At higher frequencies, we observe bands located at around 250, 400, 500, 550, and 680 cm$^{-1}$.
A comparison with Raman data reported for similar perovskites suggest that these bands are related to different oxygen motions.\cite{ISH04, SOU03}

For the brownmillerite ($x=1$), the Raman spectrum is significantly different from those of the perovskites for $x=0-0.75$. 
In comparison to the spectra for $x=0-0.75$ the $x=1$ spectrum, which agrees well with that already reported of the same material\cite{KAK04}, contains more bands, especially in the low-frequency part below 300 cm$^{-1}$, and, in general, these are stronger and more sharper in nature. 
The presence of several strong bands in the Raman spectrum reflects the presence of many more distinct crystallographic sites for the ions in the material.

The Raman spectra of the hydrated materials are shown in Fig.~\ref{Ramanhydrated}.
For $x=0.10-0.75$ the spectra of the dry and hydrated equivalents are generally similar, and only a few differences can be observed.
In the low-frequency part, the strongest effect concerns the  70--90 cm$^{-1}$ band, assigned to deformational motions of Ba-[(Zr/In)O$_{6}$] units, which is up-shifted and decreases in intensity, while the 130 cm$^{-1}$ band, related to Ba-[(Zr/In)O$_{6}$] stretches, in contrast is basically not changed at all.
The fact that the Ba-[(Zr/In)O$_{6}$] stretch band is not affected by the hydration is linked to that it  mainly involves motion of the Ba$^{2+}$ ions, \textit{i.e.} the changes in the oxygen sublattice do not influence this band.
The effects of the changes in the oxygen sublattice upon hydration are instead revealed by the oxygen bands, which are found in the 200--900 cm$^{-1}$ range.
As seen in the figure, there is an intensity increase of the 250 cm$^{-1}$ band and a decrease in intensity of the 680 cm$^{-1}$ band.
Moreover, the 500 cm$^{-1}$ band, which is only present in the spectrum for the dry $x=0.75$ material, disappears upon hydration.
For $x=1$, the effect of hydration is strong.
As seen in the figure, the strong oxygen band at around 600 cm$^{-1}$ is here absent and at lower frequencies fewer bands than in the spectrum of the dry material are observed. 

\section{Discussion}

As pointed out in Sec.~\ref{theory}, the IR spectrum of a perfectly cubic perovskite, ABO$_{3}$, should contain three relatively narrow bands, related to A-[BO$_{6}$] stretches, O-B-O bends and B-O stretches, respectively, while the Raman spectrum should ideally be featureless.\cite{LAS57}
As seen in Figs.~\ref{IR}--\ref{Ramanhydrated} this is not the case for any of the compositions in the BaIn$_{x}$Zr$_{1-x}$O$_{3-x/2}$ (\textit{x}=0--0.75) series.
This suggests that although all of the investigated perovskites are found to be cubic from an X-ray and neutron diffraction point of view the cubic symmetry must be locally distorted.
Perhaps surprisingly, even for the undoped material, BaZrO$_{3}$, with neither dopant atoms nor oxygen vacancies in the structure, we find broad bands with shoulders in the IR spectrum and the presence of several bands in the Raman spectrum, indicative of local structural distortions of the cubic structure. 
One should note that the Raman spectrum of BaZrO$_{3}$ has previously been suggested to come from second order scattering.\cite{CHA99,GEN97}
However, both the presence of the 200 cm$^{-1}$ band, related to a torsional motion of the unit cell, as well as the shoulders of the $\nu_3$ band in the IR spectrum of the same material (Fig.~\ref{IR}), indicates that the activation of Raman bands is the result of local structural distortions.

Considering first the spectra of the dry materials, the gradual broadening and appearance of new bands that grow in intensity with increasing In concentration from $x=0.10$ to $x=0.75$ in both the IR and Raman spectra show that the local structural distortions become successively more pronounced.
Upon further increasing the In concentration up to $x=1$, the vibrational spectra change significantly as the structural distortions are no longer local in nature and instead a new long-range ordered brownmillerite structure is adopted. 
Specifically, both the IR and the Raman spectra now contain more bands and the Raman spectrum in particular is more intense.
The presence of more bands in the spectra of the brownmillerite ($x=1$) is expected as discussed in Sec.~\ref{theory}.

The substitution of Zr atoms for In atoms evidently leads to local distortions of the crystal structures for $x=0.10-0.75$.
Given the size and charge difference between Zr$^{4+}$ (0.72 {\AA}) and In$^{3+}$ (0.80 {\AA}) it is not surprising that some local structural rearrangement is necessary to accommodate the larger dopant. 
Both the IR spectra in Fig.~\ref{IR} and the Raman spectra in Fig.~\ref{Ramandry} reveals the activation of both low-frequency modes, $<$~200~cm$^{-1}$, attributed to deformational motions of the Ba-[(Zr/In)O$_6$] units, and high-frequency oxygen modes between 200 and 800 cm$^{-1}$.
These results provide strong evidence for displacements of the cations and the oxygens away from their high-symmetry crystallographic sites expected in a cubic structure. 
The presence of static B site and oxygen ion disorder in the investigated materials is supported by the large atomic displacement parameters (ADPs) obtained from refined diffraction data taken at 10 K for as-prepared BaIn$_{x}$Zr$_{1-x}$O$_{3-x/2}$ ($x=0.25$ and 0.50).\cite{AHM06_3}
In particular, the ADPs of the B site and the oxygen ions increase gradually with the In concentration in agreement with our results.\cite{AHM06_3,AHM06_4}
Importantly, the structural distortion appears to be largely uncorrelated to the oxygen vacancies since both the IR and Raman spectra of the dry (many oxygen vacancies) and hydrated (few oxygen vacancies) materials look essentially the same, see Figs.~\ref{IR} and \ref{Ramanhydrated}. 

Further insight into the likely nature of the local structural distortions that accompany the introduction of the dopant atoms, in particular with relation to the oxygen ions, can be obtained from the comparison of the structures of the end members, \textit{i.e.} the BaZrO$_{3}$ perovskite structure and the Ba$_2$In$_2$O$_5$ brownmillerite structure, as shown in Fig.~\ref{structure}.
Here, the oxygen complete perovskite layers in Ba$_2$In$_2$O$_5$ exhibit correlated tilts of the InO$_{6}$ octahedra that require displacement of the oxygen ions from the basal plane.
This tilting reflects the structure reducing bond strain from the competing Ba-O and In-O interactions, in an analogous manner to that observed for perovskite systems accommodating large B site ions, \textit{e.g.} BaCeO$_{3}$\cite{JAC71}.
In general a larger B site ion separation is expected with increasing In concentration, and 
this is reflected by the expansion of the cubic lattice parameter for the dry BaIn$_{x}$Zr$_{1-x}$O$_{3-x/2}$ ($x=0-0.75$) samples.\cite{AHM06_2}
However, the magnitude of lattice expansion is considerable smaller than would be expected from a linear dependence reflecting a simple addition of the average ionic radii of the B site cations and the oxide ion. 
This further supports the presence of localized tilts of the InO$_{5}$ units in the dry In-doped BaZrO$_3$ samples which relax the B-O-B bond angle from being linear, thus allowing shorter B site separations and concomitantly leads to the activation of otherwise Raman- and IR-inactive modes.  

Given a random distribution of the dopant atoms it is evident that the effect will not be limited to the immediate neighborhood of the dopant atom, but already for relatively low In concentrations structural distortions will distribute throughout the perovskite structure.
Based on the Raman spectra in Fig.~\ref{Ramandry} a threshold for the distribution of structural distortions throughout the entire perovskite structure may lie in between $x=0.10$ and $x=0.25$, where the Raman bands get significantly stronger.
The gradual increase in intensity of the shoulders of the IR bands and the bands in the Raman spectra for frequencies $\gtrsim$~200 cm$^{-1}$ with increasing In concentration from $x=0.25$ to $x=0.75$ then most likely reflects an increase of the average tilt angle  for the (In/Zr)O$_{6}$ octahedra.

In a previous work on the BaIn$_{x}$Zr$_{1-x}$O$_{3-x/2}$ ($x~=~0.25-0.75$) system we showed that an increased In concentration leads to a stronger tendency for strong hydrogen bond formation between a proton and a neighboring oxygen as a result of more protons in non-symmetrical configurations, such as close to dopant atoms (Zr-OH-In) and/or oxygen vacancies (Zr-OH-In-Vacancy).\cite{KAR05} 
As the migration of protons through the structure requires breaking and formation of hydrogen bonds, oxygen vacancies and dopant atoms in the structure influences the proton mobility.
However, based on our result in the present work, we believe that the proton mobility is not solely affected by the presence of such defects but is also affected by the local distortions of the cubic structure.

Table~\ref{conductivity} compiles reported bulk proton conductivities for $x=0.10$, 0.25 and 0.75.
The table shows that the proton conductivity increases roughly by a factor of 9 between $x=0.10$ and $x=0.25$, and by a factor of 3.5 between $x=0.25$ and $x=0.75$.
\begin{table}
\caption{\label{conductivity}
Literature data over the proton conductivities in the BaIn$_{x}$Zr$_{1-x}$O$_{3-x/2}$ ($x=0-0.75$) series, measured at 300$^{\circ}$C. }
\begin{ruledtabular}
\begin{tabular}{lcc}
Composition & Proton conductivity /Scm$^{-1}$ & Reference \\
\hline
$x=0.10$ & 2.9$\cdot$10$^{-5}$ & \cite{KRE01}\footnote{The proton conductivity of the $x=0.10$ material has been calculated from the reported data of the proton mobility (1.26$\cdot$10$^{-8}$ cm$^{2}$s$^{-1}$ at 300$^{\circ}$C) and a hydration degree of 55{\%}, using the Nernst-Einstein equation\cite{SLA91}.}\\
$x=0.25$ & 2.6$\cdot$10$^{-4}$ & \cite{AHM06_3} \\
$x=0.75$ & 8.9$\cdot$10$^{-4}$ & \cite{AHM06_2} \\
\end{tabular}
\end{ruledtabular}
\end{table}
Evidently, the proton conductivity scales well with the concentration of protons in the materials between $x=0.25$ and $x=0.75$, for which the degree of hydration is practically the same\cite{KAR07_Tosca}.
This implies that the increase in proton conductivity between $x=0.25$ and $x=0.75$ is, at least predominantly, governed by the increased concentration of protons in the material.
Consequently, at these dopant levels, the average mobility of protons in the perovskite structure is largely the same and the gradually increasing magnitude of local structural distortions with In concentration does not significantly affect the proton mobility.
Although initially this may seem surprising, this agrees with the fact that the dopant induced structural distortions are likely to have a critical threshold in between $x=0.10-0.25$, above which the tilting of the (In/Zr)O$_{6}$ units has spread throughout the material.
Interestingly, this compositional threshold compares well to the rise in conductivity by almost one decade between $x=0.10$ and $x=0.25$.
This indicates that the presence of extended structural distortions (although with no long-range coherence in terms of the length scales probed by diffraction) of the cubic structure, obtained for $x>0.10$, may indeed be favorable for fast proton transport.

\section{Conclusions}

By using Raman and infrared spectroscopies we have investigated the effect of dopant level and hydration on the short-range structure of the proton conducting perovskite system BaIn$_{x}$Zr$_{1-x}$O$_{3-x/2}$ ($x=0-0.75$).
The results show that the doping leads to significant local distortions of the average cubic structure of these materials. 

By gradually increasing the In concentration from $x=0$ to $x=0.75$ new bands appear and grow in intensity in both the IR and Raman spectra, showing that the structure becomes successively more and more locally distorted.
Importantly, the local structural distortions are largely uncorrelated to the presence of oxygen vacancies in the structure, but instead mainly driven by the size difference between the In$^{3+}$ and Zr$^{4+}$ ions, which leads to tilting of the (In/Zr)O$_{6}$ octahedra.
Based on our results, we suggest that there is a threshold between $x=0.10$ and $x=0.25$, above which the dopant induced structural distortions have spread throughout the perovskite structure.

By comparing our spectroscopic results to reported proton conductivities of the investigated compositions in the BaIn$_{x}$Zr$_{1-x}$O$_{3-x/2}$ ($x=0-0.75$) system, we suggest that the presence of an extended structural distortion, established between $x=0.10$ and $x=0.25$, may be favorable for fast proton transport whereas at higher dopant levels the long-range proton mobility remains largely unchanged.

\acknowledgments{
This work was financially supported by the Swedish National Graduate School in Materials Science and the Swedish Research Council.}


\end{document}